\documentclass[twocolumn,preprintnumbers,superscriptaddress,endnote,nofootinbib,prl]{revtex4}
\usepackage{graphicx}
\usepackage{amsmath}
\usepackage[toc]{appendix}
\newcommand{\vev}[1]{\langle {#1} \rangle}
\newcommand{\lsim}{\lesssim}
\newcommand{\gsim}{\gtrsim}

\newcommand{\eq}[1]{Eq.~(\ref{#1})}

\newcommand{\ord}[1]{\mathcal{O}{(#1)}}
\newcommand{\beq}{\begin{equation}}
\newcommand{\eeq}{\end{equation}}
\newcommand{\bea}{\begin{eqnarray}}
\newcommand{\eea}{\end{eqnarray}}
\newcommand{\eps}{\varepsilon}

\begin{document}

\pagestyle{plain}

\title{\boldmath Implications of a Light ``Dark Higgs" Solution to the
$g_\mu-2$ Discrepancy}

\author{Chien-Yi Chen
\footnote{email: cychen@uvic.ca}
}
\affiliation{ Department of Physics, Brookhaven National Laboratory, Upton, NY 11973, USA}
\affiliation{Department of Physics and Astronomy, University of Victoria, Victoria, BC V8P 5C2, Canada}
\affiliation{Perimeter Institute for Theoretical Physics, Waterloo, ON N2J 2W9, Canada}

\author{Hooman Davoudiasl
\footnote{email: hooman@bnl.gov}
}
\affiliation{ Department of Physics, Brookhaven National Laboratory, Upton, NY 11973, USA}

\author{William J. Marciano
\footnote{email: marciano@bnl.gov}
}
\affiliation{ Department of Physics, Brookhaven National Laboratory, Upton, NY 11973, USA}

\author{Cen Zhang
\footnote{email: cenzhang@bnl.gov}
}
\affiliation{ Department of Physics, Brookhaven National Laboratory, Upton, NY 11973, USA}

\begin{abstract}

A light scalar $\phi$ with mass $\lsim 1$~GeV and muonic coupling
$\mathcal{O}(10^{-3})$ would explain the 3.5 $\sigma$ discrepancy between the
Standard Model (SM) muon $g-2$ prediction and experiment.  Such a scalar can be
associated with a light remnant of the Higgs mechanism in the``dark" sector.  
We suggest $\phi\to l^+l^-$ bump hunting in $\mu\to
e\nu\bar\nu\phi$, $\mu^-p\to\nu_\mu n\phi$ (muon capture), and $K^\pm\to
\mu^\pm\nu\phi$ decays as direct probes of this scenario.  In a general setup, 
a potentially observable muon electric dipole moment $\lsim 10^{-23}\ e \cdot\textrm{cm}$ and lepton flavor
violating decays $\tau\to\mu
(e) \phi$ or $\mu \to e \phi$ can also arise.  Depending on parameters, 
a deviation in BR($H\to\mu^+\mu^-$) from
SM expectations, due to Higgs coupling misalignment, can result.  We illustrate how the
requisite interactions can be mediated by weak scale
vector-like leptons that typically lie within the reach of future LHC
measurements.

\end{abstract} \maketitle
\section{Introduction}
The well-established existence of cosmic dark matter (DM) - a form of matter
that does not significantly interact with ordinary atoms - furnishes us with
clear evidence for physics beyond the Standard Model (SM).  In many models, such
as supersymmetry, DM naturally fits in extensions of the electroweak sector
that attempt to explain properties of the Higgs potential.  However,
more generally, the
dominance of cosmic DM over visible matter could argue for an entirely new
sector of particle physics - the ``dark sector" - endowed with its own forces
and particles, largely decoupled from the SM \cite{Essig:2013lka}.  The dark
sector might only have faint interactions with our visible sector, mediated by
the so called portal
\cite{Patt:2006fw,Davoudiasl:2004be,McKeen:2012av,Batell:2009zp}
states that reside in both worlds.

In this work, we examine the possibility that a SM singlet light scalar $\phi$
residing primarily in the dark sector can account for the long-standing 3.5 $\sigma$
discrepancy between the SM prediction and measured value of the muon anomalous
magnetic moment $a_\mu=(g_\mu-2)/2$,
\begin{equation}
	\Delta a_\mu\equiv a_\mu^\text{exp}-a_\mu^\text{SM}=276(80)\times10^{-11}
	\,,
	\label{eq1}
\end{equation}
which we have updated to include NNLO hadronic vacuum polarization effects
\cite{Agashe:2014kda,Kurz:2014wya}.  We consider scalar masses $m_\phi \lsim
1$~GeV, in the framework of a ``dark photon" \cite{Davoudiasl:2012ag} scenario
with a simple ultraviolet (UV) completion, {\it i.e.} dark weak scale
vector-like leptons and one extra Higgs doublet with hypercharge.  All new
particles (modulo the dark photon) carry a dark $U(1)_d$ charge, leading, via
mixing, to a low energy theory with the muonic couplings to $\phi$ necessary to
explain Eq.~(\ref{eq1}).  In our scenario, $\phi$ is associated with $U(1)_d$
breaking in the dark sector, {\it i.e.} it is a ``dark Higgs'' remnant of
``dark'' symmetry breaking.

A dark sector $U(1)_d$ force, with an associated dark vector boson $\gamma_d$
$\lsim$ GeV-scale, has been motivated for some time from various astrophysical
signals ascribed to DM \cite{ArkaniHamed:2008qn}.  A ``dark Higgs mechanism"
can be invoked as a primary source of $\gamma_d$ mass.  Kinetic mixing
\cite{Holdom:1985ag} between $U(1)_d$ and $U(1)_Y$ of hypercharge can allow
$\gamma_d$ to couple to the SM electromagnetic current, where $\gamma_d$ is
then often referred to as a ``dark photon''. If the kinetic mixing is
sufficient, $\gamma_d$ may itself play an important role in explaining
$g_\mu-2$ \cite{Pospelov:2008zw}; however, we do not consider that possibility
here. Instead, we assume the $\phi$ is responsible for the bulk of the
discrepancy.

The dark photon model can be generalized by assuming that $\gamma_d$ and the SM
$Z$ boson couple to a dark second Higgs doublet that induces $\gamma_d-Z$
mass-mixing, in which case the resulting light $Z_d$ (a linear $\gamma_d-Z$
combination) acts much like a light ``dark $Z$" \cite{Davoudiasl:2012ag}, with
interesting additional implications, such as changes in the low-$q^2$ running
of the weak mixing angle \cite{Davoudiasl:2014kua,Davoudiasl:2015bua}, and rare
decays of $K$, $B$ and $H$ particles into final states with $Z_d$'s
\cite{Davoudiasl:2012ag}.

The aforementioned kinetic mixing can naturally arise
in the dark photon scenario from quantum loops of
heavy vector fermions that carry both $U(1)_d$ and $U(1)_Y$ charges
\cite{Holdom:1985ag,Davoudiasl:2012ig}.  In
principle, such fermions could occur near the weak scale $\sim$ 250 GeV,
especially if they play a role in electroweak symmetry breaking.  As precision
electroweak and collider bounds generally disfavor states that carry $SU(2)_L$
or color $SU(3)$ charges, one may assume for illustration
that in its simplest version the
lightest vector-like fermions have the quantum numbers of the SM right-handed
charged leptons.  
Therefore, on general grounds, vector-like leptons, as well as SM singlet and
doublet scalars that carry $U(1)_d$ charges are well-motivated ingredients
underlying the dark $Z$ model \cite{Davoudiasl:2012ag}.

A direct low energy probe of our framework is $\phi$ bump hunting in
$\mu^\pm\to e^\pm\nu\bar\nu\phi$, or $\mu^-p\to\nu_\mu n\phi$ (muon capture
on nuclei),
and $K^\pm \to \mu^\pm \nu_\mu \phi$, with $\phi$ decaying into lepton pairs or
invisibly to light dark particles after being radiated by the muon.
Typically, we may also expect new sources of CP and lepton flavor violations to
arise in our scenario due to mass-scalar coupling misalignment, leading to a
potentially detectable muon {\it electric dipole moment} $\ord{10^{-23} e
\,\rm{cm}}$ and leptonic decays $\ell \to \ell' \phi$, respectively. In the
appendix, we provide a simple high scale model that can typically accommodate 
dominant lepton flavor diagonal couplings with sufficiently suppressed flavor
violating ones for $\phi$ to evade experimental constraints, some of which we
later discuss.  This model can also support a realistic neutrino mass
matrix, as briefly discussed in the Appendix.

An important ingredient of the above setup is that dark and visible sector Higgs
interactions together yield a new source of SM charged lepton masses.  Hence, depending 
on the parameters in the dark sector, one could, in addition to $\phi$
effects, also expect departures in the 125~GeV Higgs branching fractions into
$e, \mu, \tau$ pairs from SM expectations. Such deviations may be measurable at
the LHC in the coming years as Higgs decay statistics continue to improve.

\section{Lepton Dipole Moments}

In this work, the main motivation for introducing 
the scalar $\phi$ is its potential role as a new contribution to $g_\mu-2$.
Furthermore, we will also assume that $\phi$ can contribute to both the
magnetic and electric dipole moments of leptons, through CP-conserving and
violating couplings.

In general, the flavor diagonal Yukawa couplings can be parametrized
(relative to the real CP conserving charged lepton mass matrix) as

\begin{figure}[bth]
	\begin{center}
		\includegraphics[width=.25\linewidth]{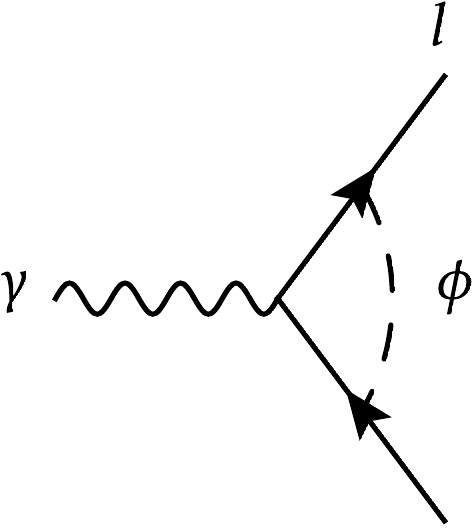}
	\end{center}
	\caption{One loop $\phi$ contribution to lepton dipole moments.}
	\label{fig:dipolediagram}
\end{figure}
\begin{figure}[bth]
	\begin{center}
		\includegraphics[width=.99\linewidth]{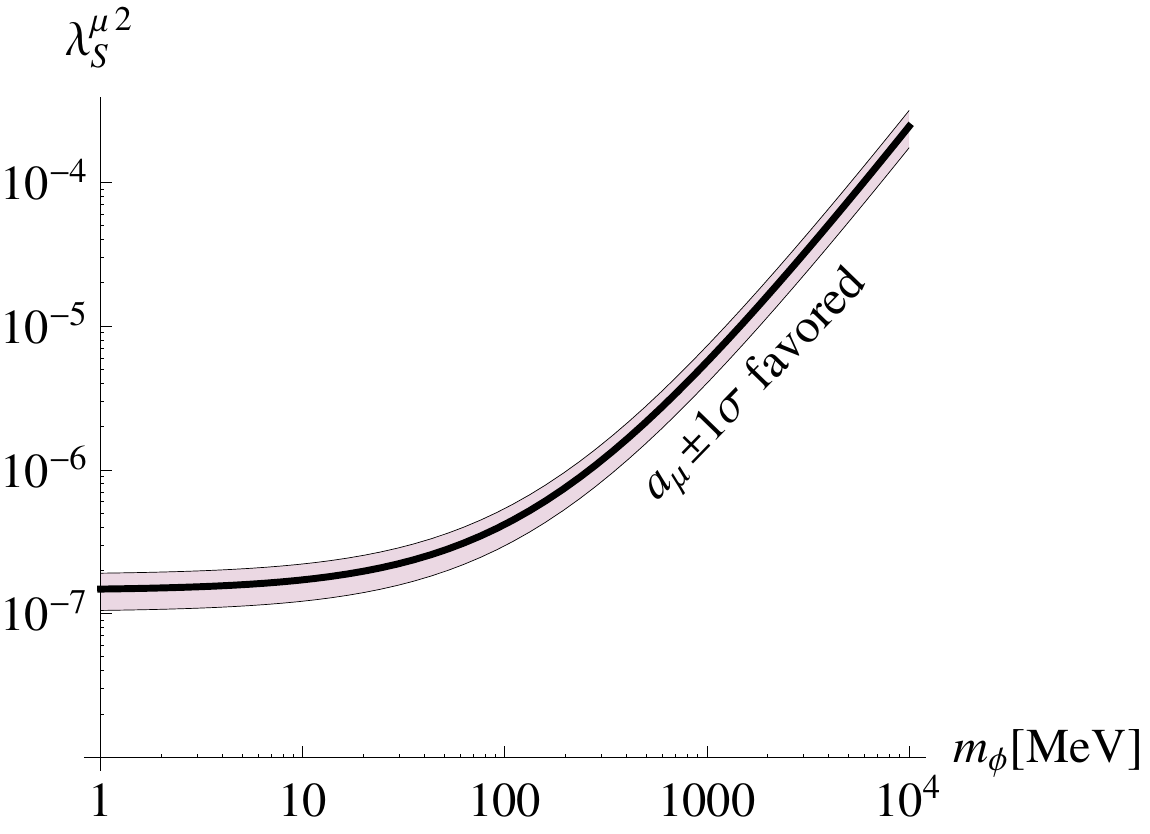}
	\end{center}
	\caption{Central values and one sigma band of $\lambda_{S}^\mu$,
	required by the measured value of $\Delta a_\mu$ in Eq.~(\ref{eq1}).}
	\label{fig:lambdaplot}
\end{figure}


\begin{align}
	\mathcal{L}_{\phi \ell\ell}=
	- \phi\bar\ell\left( \lambda_S^\ell+i\lambda_P^\ell\gamma_5 \right)\ell
	\label{Lphill}
\end{align}
where $\ell=e,\mu,\tau$, and $\lambda_S^\ell$ ($\lambda_P^\ell$) is the CP-even (odd)
dark Yukawa coupling.  At one-loop level these couplings induce additional
contributions to the dipole moments of leptons, as shown in
Figure~\ref{fig:dipolediagram}.  We find they imply
\cite{Leveille:1977rc,TuckerSmith:2010ra}
\begin{flalign}
	\Delta a_\ell=& \frac{{\lambda_S^\ell}^2}{8 \pi ^2}r^{-2}
	\int^1_0\mathrm{d}z\frac{\left( 1+z \right)\left( 1-z \right)^2}
	{r^{-2}\left( 1-z \right)^2+z}
	\nonumber\\
	&-\frac{{\lambda_P^\ell}^2}{8 \pi ^2}r^{-2}
	\int^1_0\mathrm{d}z\frac{\left( 1-z \right)^3}
	{r^{-2}\left( 1-z \right)^2+z}
\label{Dela}
\end{flalign}
and for the lepton electric dipole moment 
\begin{flalign}
	d_\ell=\frac{\lambda_S^\ell\lambda_P^\ell}{4\pi^2}\frac{e}{2m_\ell}
	r^{-2} \int^1_0\mathrm{d}z\frac{\left( 1-z \right)^2}
	{r^{-2}\left( 1-z \right)^2+z}
\label{d}
\end{flalign}
where $r=m_\phi/m_\ell$.  We present analytic expressions 
for these integrals in the Appendix.  In the limit $r\to 0$ ({\it i.e.}, light $\phi$)
we have
\begin{flalign}
	\Delta a_\ell&=\frac{1}{16 \pi ^2}
	\left(3 {\lambda_S^\ell}^2 - {\lambda_P^\ell}^2\right)
	\\
	d_\ell&=\frac{\lambda_S^\ell\lambda_P^\ell}{4 \pi ^2
	}\frac{e}{2m_\ell}\,.
\end{flalign}
In Figure~\ref{fig:lambdaplot} we illustrate (ignoring $\lambda_P^\mu$) the
region of ${\lambda_S^\mu}^2$,
$m_\phi$ favored by Eq.~(\ref{eq1}) with one-sigma uncertainty. 

In the electron case, there is no significant deviation from the SM $g_e-2$ prediction.  
However, if $\Delta a_\mu$ is taken to be $276\times10^{-11}$
and we assume $\lambda_S^e \sim\frac{m_e}{m_\mu}\lambda_S^\mu$, (for negligible
$\lambda_P^e$ effects), we find that $|\Delta a_e|<10^{-13}$ for all $m_\phi$,
well below the current experimental constraint $\Delta
a_e=\left(-0.91\pm0.82\right)\times10^{-12}$ \cite{Aoyama:2014sxa}. Hence,  $\Delta
a_e$ consistent with zero is easily
accommodated in our scenario for reasonable couplings.

The ratio between electric and anomalous magnetic moments in the $r\to0$ limit
is
\begin{flalign}
	\frac{d_\ell}{\Delta a_\ell}=&\frac{e}{2m_\ell}
	\frac{4\tan\theta_\ell}{3-\tan^2\theta_\ell}
	\approx \frac{e}{2m_\ell}\frac{4}{3}\tan\theta_\ell
\end{flalign}
where we define $\tan\theta_\ell=\lambda_P^\ell/\lambda_S^\ell$.  
Note that under the opposite $r\to\infty$ limit both $\Delta a_\ell$ and
$d_\ell$ vanish (for earlier related work, see Ref.~\cite{GengandNg}) .
In principle one should also add the two-loop Barr-Zee contribution
\cite{Barr:1990vd} to $d_\ell$. However, for the muon, we expect it to be subdominant. 
For the electron it is potentially more important.

\begin{figure}[bth]
	\begin{center}
		\includegraphics[width=.99\linewidth]{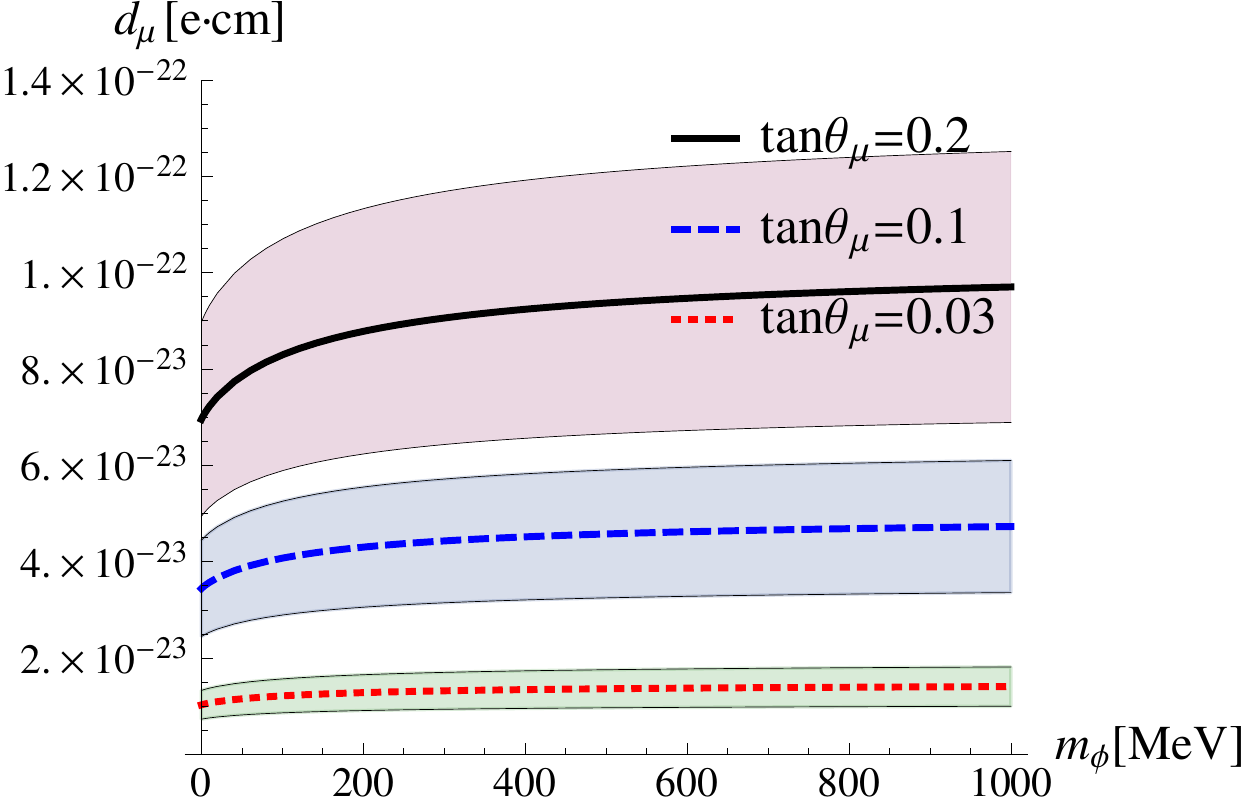}
	\end{center}
	\caption{Muon electric dipole moment, for various $CP$-violating phases,
		assuming that $\Delta a_\mu$ agrees with the measured value of
	$g_\mu-2$ within one sigma.}
	\label{fig:edmplot}
\end{figure}

The one-loop induced electric dipole moment of a lepton can be written as
$d_\ell=2.36\times10^{-15}\lambda_S^\ell\lambda_P^\ell (m_\mu/m_\ell)
e\cdot\mathrm{cm}$.  To estimate the size of the muon electric dipole moment,
we assume that the
$g_\mu-2$ central anomaly can be solely explained by the scalar contribution to
$a_\mu$ from Eq.~(\ref{Dela}).  (The required ${\lambda_s^\mu}^2$ central values as
a function of $m_\phi$ are given in Figure~\ref{fig:lambdaplot} with a
one-sigma spread.)  This will determine the dark Yukawa couplings
$\lambda_{S}^\mu$, up to the CP-violating phase $\theta_\mu$.  For any given
value of $\tan\theta_\mu$, we can compute $d_\mu$ as a function of the $\phi$
mass.  Results are shown in Figure~\ref{fig:edmplot}, for $\tan\theta_\mu=0.2$,
$0.1$, and $0.03$.  We see that for reasonable values of $\tan\theta_\mu$, the
muon electric dipole moment can reach about
$10^{-22}\sim10^{-23}e\cdot\mathrm{cm}$.  That is to be compared with the
current bound $|d_\mu|<1.8\times10^{-19}e\cdot \mathrm{cm}$
\cite{Bennett:2008dy}.  Possible muon storage ring measurements of $d_\mu$ with
sensitivity $10^{-24}\sim10^{-25}e\cdot\mathrm{cm}$ have been envisioned, but
for now none are planned \cite{Semertzidis:2003iq,Semertzidis:1999kv}.
In principle, they could explore down to $\tan\theta_\mu\sim0.0003$ in our
scenario.

It is possible that eq.~(\ref{d}) could also lead to a detectable electric dipole
moment for the electron.  Intuitively, we might expect that
$\lambda_{S,P}^\ell\propto m_\ell$, i.e.~proportional to the relative chiral
symmetry breaking mass scale, even though this is model dependent.
Assuming this relation we expect that $d_\mu/d_e$ may be of order
$10^{5}\sim10^{6}$ for $m_\phi$ within the range $[10,1000]$ MeV.  That means
$d_e$ could turn out to be $10^{-28}\sim10^{-29}e\cdot$cm, which is to
be compared with the current bound \cite{Baron:2013eja}
\begin{equation}
	|d_e|<8.7\times10^{-29}e\cdot\mathrm{cm}
	\label{eq:debound}
\end{equation}
Hence $d_e$ could potentially be within the reach of future experiments which are
expected to probe down to $|d_e|\sim\mathcal{O}(10^{-30})~e\cdot$cm.

\section{Direct Signals in Rare Lepton Flavor Preserving Processes}

A direct consequence of our solution to $g_\mu-2$ is the possibility of $\phi$
emission in rare lepton flavor preserving processes involving initial or final state muons. 
In what follows, we will consider muon and kaon interactions that could offer
promising search avenues for our scenario.

\begin{figure}[bth]
	\begin{center}
		\includegraphics[width=.99\linewidth]{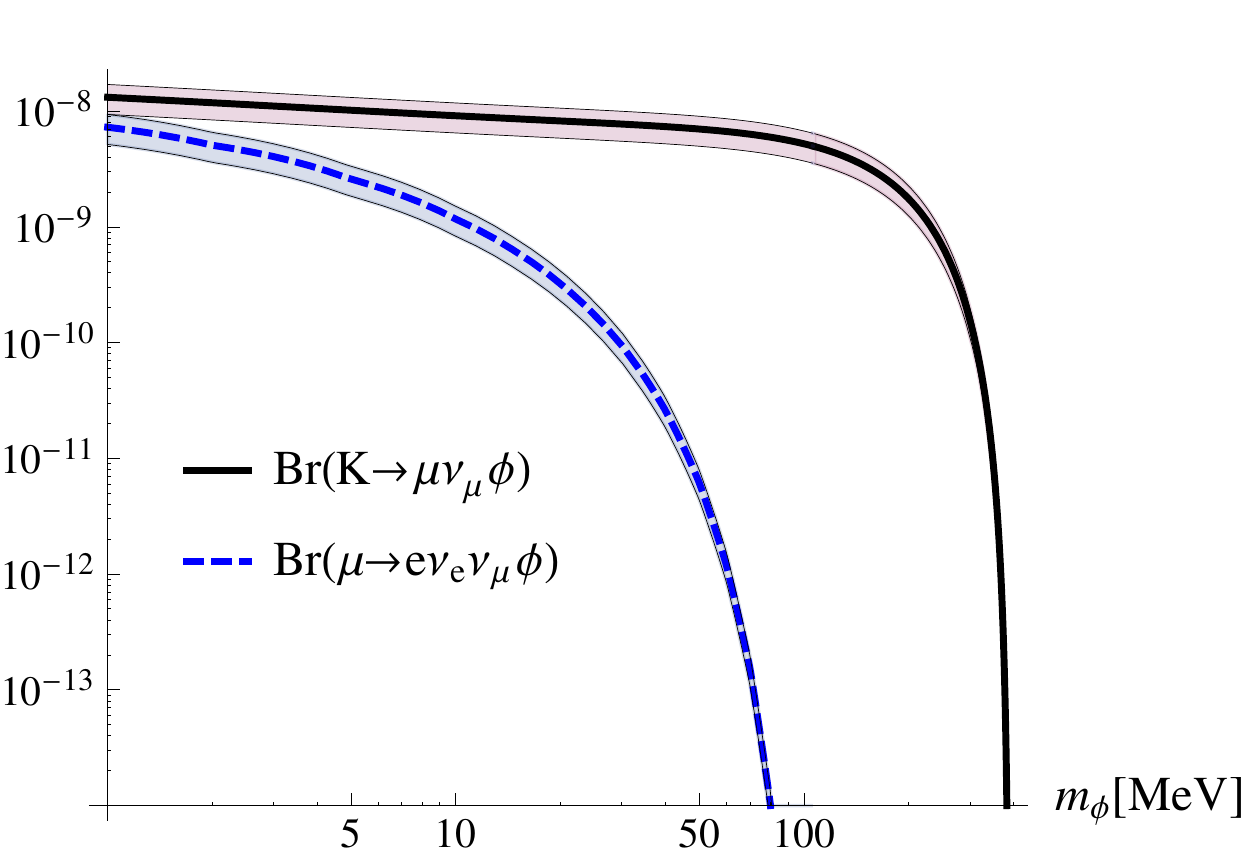}
	\end{center}
	\caption{Branching ratios of $\mu\to e\nu\bar\nu\phi$ and $K^+ \to
	\mu^+\nu\phi$, assuming that $\Delta a_\mu$ agrees with the measured
value of $g_\mu-2$ within one sigma.} \label{fig:brplot}
\end{figure}

{\it Muon decay:} 
We first consider $\mu \to e\, \phi\, \bar \nu_e\, \nu_\mu$, whose branching
ratio is given in Figure~\ref{fig:brplot}, for $\lambda_S$ couplings that
accommodate $g_\mu-2$.  This $e\, \phi + \text{``invisible"}$ signal can be
probed with intense muon sources such as Mu3e; see for
example Ref.~\cite{Echenard:2014lma} for a discussion based on the similar case
of dark photons.  Here, assuming $m_\phi \lsim m_\mu$ and an $\ord{1}$ branching
fraction for $\phi\to e^+ e^-$, we may expect sensitivity to $\lambda_S^\mu$
similar to that for a dark photon with kinetic mixing parameter $\eps$, where $\eps e\to
\lambda^\mu_S$.  We note
that while the presence of the $e^+e^-$ mode is not strictly required in our
scenario, the assumed muon coupling does imply a nonzero loop-induced
branching fraction for $\phi\to \gamma\gamma$, which may not have detection
prospects similar to that of the $e^+ e^-$ final state, depending on the
experimental setup (such as the use of a nonzero magnetic field for event
selection).  While the current bounds are not very constraining for our
scenario, future measurements, such as those discussed in
Ref.~\cite{Echenard:2014lma} can potentially probe $\lambda_S^\mu \lsim
10^{-4}$ for $m_\phi\sim20-80$ MeV, in the case of $\phi \to e^+e^-$ dominance,
which would cover much of the parameter space relevant for $m_\phi \lsim
m_\mu$ that resolves the $g_\mu-2$ discrepancy.

{\it Muon capture:} 
The decay $\mu^-\to e^-\nu\bar\nu\phi$ with $\phi\to e^+e^-$ can also be
searched for in bound muon decay studies such as Mu2e at Fermilab and Comet at
J-Parc where more than $10^{17}$ muons are expected to be stopped
in an $Al$ target where they form $\mu^-Al$ atoms.  About half of those stopped
muons undergo ordinary muon decay $\mu\to e\nu\bar\nu$ in orbit while the other
half undergo capture $\mu^-Al\to\nu_\mu\ M\!g$.  The capture process and decay in
orbit are both potential $\phi$ sources.  Using either mode to search for
$\phi$'s, decay or capture, represents an interesting extension of the
muon conversion experiments getting underway.  Their viability will depend on
reconfiguring the detectors to observe $e^+e^-$ pairs from $\phi$ decays with
invariant mass $m_\phi$ above background.

{\it Kaon decay:} The scalar $\phi$ can, in principle, also be radiated from muons in 
$K^+ \to \mu^+ \nu_\mu$ decays, via the $\phi$-muon coupling~\cite{Barger:2011mt,Beranek:2012ey,Carlson:2013mya}.  The
branching ratio is given in
Figure~\ref{fig:brplot}.  The rate is more than an order of magnitude below the current bound
\cite{Agashe:2014kda}
\begin{flalign}
	BR(K^+\to\mu\nu\phi,\ \phi\to\mu^+\mu^-)<4.1\times10^{-7}.
\end{flalign}
However, the ongoing NA62 experiment at CERN may be potentially sensitive to
this decay mode for $m_\phi\sim 200 - 300$ MeV. 

For $m_{\phi} < 200$ MeV, the decay $K^+\to \mu\nu\phi, \phi\to e^+e^-$ may be observable over the SM background $K^+\to \mu\nu e^+e^-$, which has been measured by experiment E865 at Brookhaven~\cite{Poblaguev:2002ug}
\begin{flalign}
	BR(K^+\to\mu\nu e^+e^-&) = 7.06\pm 0.16\pm0.26\times10^{-8} \\ \nonumber
	 &\rm{for}\;\; M_{e^+e^-}>145 \;\;\rm{MeV}.
\end{flalign}
Although a binned search for narrow resonances in the $e^+e^-$ invariant mass spectrum was not carried out in that data analysis, given the large number of $K^+$ decays analyzed $\sim 3\times 10^{10}$,
a $BR(K^+\to\mu\nu\phi, \phi\to e^+e^-)$ sensitivity of $10^{-8} -10^{-9}$ may be possible over an interesting range of $\phi$ masses 
($m_\phi \sim 145-200$ MeV) with existing data. That would provide a test of the model under the assumption $\phi \to e^+e^-$ is the dominant decay mode of the $\phi$. More detailed analyses can be found in Ref.~\cite{lange}.
 
{\it Pion decay:}
For light $\phi \lsim 30$ MeV, one can look for the decay $\pi^+\to
\mu^+\nu_\mu\phi$, $\phi\to e^+e^-$ by searching for a $e^+e^-$ bump at high-intensity 
charged pion sources such as NA62 or beam-dump experiments.
Exploring that possibility is interesting and worthy of study.  However, addressing 
current bounds requires thorough background studies and depends on
the promptness of the decay, which is model dependent. For that reason, further
discussion is beyond the scope of this paper. 

\section{A Concrete UV Model}

Here, we provide a possible UV completion of our low energy effective theory,
which leads to the assumed coupling in Eq.~(\ref{Lphill}).  This model can also
provide the requisite ingredients for a potentially viable ``dark" $Z$ model
(see, for example, Ref.~\cite{Davoudiasl:2012ag}).   In this UV framework, all
new particles are assumed to be charged under $U(1)_d$ with the same dark
charge, unless otherwise stated, and hence we will only identify their SM
charges.  Let $X^\ell$ --- where $\ell=e,\mu,\tau$ is a flavor index ---
be vector-like fermions with the quantum numbers of right-handed SM leptons
$\ell_R$ (i.e.~$SU(2)$ singlets), and
masses $m_X^\ell \gsim \text{few}\times 100$~GeV.  We also introduce a new
Higgs scalar doublet $H_d$ and a complex scalar singlet $\phi$.  We will assume
that $H_d$ and $\phi$ have nonzero vacuum expectation values (vevs) which
spontaneously break $U(1)_d$.  As mentioned before, these ingredients can be
motivated within a dark $Z$ model \cite{Davoudiasl:2012ag}.  One can then write
down the following SM$\times U(1)_d$ invariant interactions
\begin{eqnarray}
	-\label{L1}
	{\cal L}_1 &=& m_X^{\ell \ell'} \bar X^\ell X^{\ell'} + \lambda_1 \phi \bar X^\ell_L \ell_R
	+ \lambda_2 H_d \bar L^\ell X^\ell_R \\
	&+& y_\ell H \bar L^\ell \ell_R + \mbox{\small \text{H.C.}}\,,\nonumber
\end{eqnarray}
where $L^\ell$ and $H$ refer to SM lepton and Higgs doublets, respectively.
The above interactions respect lepton flavor conservation up to soft breaking
by (small) off-diagonal masses $m_X^{\ell \ell'}$, which we will assume are the
only sources of lepton flavor violation.  In the appendix, we illustrate how the
above can be realized in a model with flavor symmetries that allow for a
realistic neutrino mass matrix.  A vacuum expectation value for $H_d$ followed by charged lepton mass matrix
diagonalization could result in misaligned $\phi$ and $H$ lepton couplings
which lead to interesting consequences, as outlined below.
In case of extension to quarks, our scenario maintains $H$ and $H_d$ alignment with the mass matrix and 
avoids quark flavor changing current constraints at the tree level.

\section{Lepton flavor violating decays}

A possible signal of our UV model is the appearance of lepton flavor violating
(LFV) interactions of the form $\lambda^{ij}_S\phi\bar l_il_j$ 
(pseudoscalar couplings are also possible, but will not be considered here).
In particular, they can give rise to $\mu \to \phi \, e$ and $\tau \to \phi \, l$, with
$l=\mu, e$.  The constraints on these interactions
depend sensitively on the dominant $\phi$ decay channels. Generally speaking,
these constraints are quite a bit weaker when $\phi \to \text{``invisible"}$ is
the dominant decay mode~\cite{Agashe:2014kda,Bayes:2014uja}\footnote{
In a scenario where the branching ratio of $\phi$ decay into invisible is 100\%, we found that 
the bounds on the off-diagonal couplings, depending on flavor, are in general 2-4 orders of magnitude weaker.}; 
we will have more comments on this case later.  Instead, let us consider the case of a visible $\phi$
with $\phi \to \mu^+\mu^-$ or $e^+ e^-$ (below the di-muon threshold).

The current upper bound on the $\mu \to 3 e$ branching fraction is
\cite{Agashe:2014kda}
\begin{equation}
	BR(\mu\to e\phi,\ \phi\to e^+e^-\ \mathrm{prompt})<10^{-12}
\end{equation}
which corresponds to limits on $|\lambda_S^{\mu e}|$
\begin{eqnarray}
\label{lfvmu}
|\lambda_S^{\mu e}| &<& 1.2\times10^{-14} \;\;\rm{for} \;\;m_\phi = 10 \;\;\rm{MeV}, \nonumber \\ 
|\lambda_S^{\mu e}| &<& 1.5\times10^{-14} \;\;\rm{for} \;\;m_\phi = 50 \;\;\rm{MeV}, \\  
|\lambda_S^{\mu e}| &<& 1.1\times10^{-13} \;\;\rm{for} \;\;m_\phi = 100 \;\;\rm{MeV}.  \nonumber
\end{eqnarray}
In the case of $\tau \to 3 l$, roughly speaking, the bounds on the
corresponding branching fractions are much weaker, $\sim  \text{few} \times
10^{-8}$~\cite{Agashe:2014kda}
\begin{flalign}
	&BR(\tau\to e\phi,\ \phi\to e^+e^-\ \mathrm{prompt})<2.7\times 10^{-8}	
	\\
	&BR(\tau\to e\phi,\ \phi\to \mu^+\mu^-\ \mathrm{prompt})<2.7\times 10^{-8}
	\\
	&BR(\tau\to \mu\phi,\ \phi\to e^+e^-\ \mathrm{prompt})<1.8\times 10^{-8}
	\\
	&BR(\tau\to \mu\phi,\ \phi\to \mu^+\mu^-\ \mathrm{prompt})<2.1\times 10^{-8}
\end{flalign}
These correspond to limits on $|\lambda_S^{\tau l}|$
\begin{eqnarray}
\label{lfvtau}
|\lambda_S^{\tau l}| &<& 1.0\times10^{-9} \;\;\rm{for} \;\;m_\phi = 50 \;\;\rm{MeV}, \nonumber \\ 
|\lambda_S^{\tau l}| &<& 1.2\times10^{-9} \;\;\rm{for} \;\;m_\phi = 500 \;\;\rm{MeV},  \\  
|\lambda_S^{\tau l}| &<& 3.5\times10^{-9} \;\;\rm{for} \;\;m_\phi = 1500 \;\;\rm{MeV}.   \nonumber
\end{eqnarray}

A rough estimate yields $m_X^{\tau l}\lsim
10$~keV (assuming vector lepton masses $m_X \sim 100$~GeV).  A simple model
of flavor, presented in the Appendix, can accommodate such a degree of LFV,
while providing Dirac masses $\sim 0.1$~eV for neutrinos.
The more constraining bound on $\mu\to e\phi$ can be taken to imply a phenomenological preference for $m_\phi\gsim 100$~MeV, so that muon decays 
to on-shell $\phi$ final states are not kinematically allowed.\footnote{Alternatively, one may consider $m_X^{\mu e} \ll
m_X^{\tau l}$, assuming for example that the $e$-flavor-breaking parameter $S^e \ll S^{\mu, \tau}$.
This would imply that $m_X^{\tau e}$ and, consequently, $\tau \to \phi\, e$ are
also suppressed, suggesting that one of the neutrinos is much lighter than the
other two (which is currently allowed by all data).}

Adhering to the types of bounds in Eqs.~(\ref{lfvmu}) and (\ref{lfvtau}) will also suppress loop induced 
LFV decays such as $\mu \to e \gamma$, $\tau \to e \gamma$ and $\tau \to \mu \gamma$.  
However, a detailed study of such effects is likely to require a 
more complete two-loop analysis~\cite{Bjorken:1977vt}, which is beyond the scope of this paper.

\section{\boldmath $H\to l^+l^-$ Misalignment}

If in addition to the SM Higgs mechanism, there exist other contributions to lepton
masses, then some misalignment between the
charged lepton mass matrix and $H\ell^+\ell^-$ couplings can also be expected.  In our framework, 
a significant source of muon mass can originate from its interactions with $\phi$, assuming that 
$\vev{\phi}$ is $\ord{100~{\rm GeV}}$.\footnote{This possibility 
can be motivated in phenomenologically viable ``dark" $Z$ models \cite{Davoudiasl:2012ag}, as a means of suppressing 
$Z$-$Z_d$ mass mixing.}   Ignoring flavor changing effects, which are interesting (especially for
$H\to\mu\tau$) but beyond the scope of this study, one can parametrize the
misalignment by a $H\ell^+\ell^-$ coupling factor relative to the SM value by
\cite{combined}
\begin{equation}
	\kappa_\ell\left( \cos\theta_\ell^H+i\gamma_5\sin\theta_\ell^H \right)
\end{equation}
where $\kappa_\ell$ scales the relative magnitude of the coupling and
$\theta_\ell^H$ allows for a $CP$-violating component.  The latter effect is
potentially very interesting for the electron, where the recent bound
on the electron electric dipole moment, as given in Eq.~(\ref{eq:debound}),
already leads to the rather prohibitive constraint
\cite{Altmannshofer:2015qra}
\begin{equation}
	|\sin\theta_e^H| < 0.017/\kappa_e\,.
\end{equation}
That sensitivity is expected to further improve by as much as two orders of
magnitude in the future as experiments probe $|d_e|\sim10^{-30}e\cdot$cm.

Recently, the $H\to\tau^+\tau^-$ decay has been measured by the ATLAS and CMS
collaborations at the LHC at better than 5 sigma.  The observed branching ratio
leads to \cite{hcouplings}
\begin{equation}
	\kappa_\tau=0.90^{+0.14}_{-0.13}
\end{equation}
consistent with the SM expectation $\kappa_\tau^{SM}=1$.  Further precision
is expected from Run II.  Measurement of $\kappa_\mu$ will be more difficult
but potentially doable in Run II of the collider if $\kappa_\mu\sim1$ or even
larger.  Run I searches for $H\to\mu^+\mu^-$ have so far been
negative, leading to the constraint \cite{hcouplings}
\begin{equation}
	\kappa_\mu=0.2^{+1.2}_{-0.2}
	\label{eq:kappamu}
\end{equation}
Although still consistent with the SM expectation $\kappa_\mu^{SM}=1$,
the central value in Eq.~(\ref{eq:kappamu}) reminds us that an enhancement or
(perhaps more likely) a suppression of $H\to\mu^+\mu^-$ is very possible.  That
would be an exciting discovery, confirming misalignment.  In the case of $H\to e^+e^-$, the SM
branching ratio of $\sim5\times10^{-9}$ is very suppressed, making that decay
mode highly unlikely to be observable unless $\kappa_e\gg1$, which would seem
to be somewhat contrived in our scenario.

\section{Additional Phenomenology}

Finally, we would like to discuss potential signals of our scenario, based
on the the UV model assumed in \eq{L1}.  To do so, we adopt somewhat specific 
values for parameters, in order to highlight some typical
possibilities for the implied general phenomenology.  As 
illustrated in the following discussion, a wide variety of possibilities can
ensue from our underlying theory and, depending on specific choices
of parameters, a number of interesting signals can arise in high energy
experiments.  A more detailed examination of such possibilities, while quite
interesting and instructive, will exceed the intended scope of our current
work.

As previously mentioned, our underlying assumption regarding
the coupling of $\phi$ to muons also suggests deviations in the Yukawa
coupling of the muon to the observed 125 GeV Higgs, because of a secondary source
for $m_\mu\simeq 106$~MeV provided via the dimension-5 operator
\beq
\lambda_1 \lambda_2 \frac{\phi \, H_d {\bar L_\mu} \mu_R}{m_X},
\label{dim5mu}
\eeq
with $m_X\equiv m_X^{\mu\mu}$, for notational simplicity.  Since $H_d$ is a
weak isodoublet, $\vev{H_d}$ contributes to electroweak symmetry breaking (EWSB).  
However, electroweak measurements currently seem to show good agreement with the SM
predictions. Hence, it is well motivated to assume that $H_d$ has a
subdominant role in EWSB: $\vev{H_d}\ll \vev{H}$.  Let us assume, for illustration,
that $\vev{H_d} \sim 30$~GeV and $m_X \sim 300$~GeV, as reasonable values.

Note that for $m_\phi$ not far from $\sim 100$~MeV, a typical value in this
work, we have $\lambda^\mu_S \simeq  \lambda_1 \lambda_2 \vev{H_d}/m_X \sim
10^{-3}$ in order to account for $g_\mu-2$ for $\lambda^\mu_P\ll
\lambda^\mu_S$ (small $\theta_\mu$).  If $\vev{\phi}\sim
100$~GeV (as in typical dark $Z$ models where such a setup can provide the requisite
suppression of $Z$-$Z_d$ mass mixing \cite{Davoudiasl:2012ag}) our choices
of parameters then imply $\lambda_1 \lambda_2 \sim 10^{-2}$. The coupling of
$\phi$ to the $X^\mu$ can also induce a large quantum loop generated scalar mass
$\delta m_\phi \sim \lambda_1 \, m_X/(4 \pi)$, which motivates the assumption
$\lambda_1 < \lambda_2$ and hence we may choose,
for example, $\lambda_2 \sim 0.2$ and $\lambda_1 \sim 0.05$.
The choice $\lambda_2 \sim 0.2$ implies a $X^\mu-\mu$ mixing
$\lambda_2\vev{H_d}/m_X\sim0.02$, which is roughly consistent with precision
bounds.

The above discussion of parameters has interesting implications for the
phenomenology of our underlying model, aspects of which we will briefly
consider.  For one thing, the coupling of the scalar doublet $H_d$ to muons
$y_\mu^d = \lambda_S \vev{\phi}/\vev{H_d} \sim 3 \times 10^{-3}$. This is
roughly a factor of $\sim 5$ larger than the muon-Higgs Yukawa coupling in the SM!
Thus, we have a scenario where the 125 GeV Higgs may have suppressed couplings
to muons, whereas the second ``dark" doublet $H_d$ may have considerably
enhanced interactions with muons.  It may, therefore, be interesting to
consider the potential resonant production of $H_d$ at a future weak scale $\mu^+\mu^-$ 
collider.  The same consideration also applies to the production of $\phi$ at
a low energy $\mu^+\mu^-$ collider, given its assumed relatively large 
coupling to muons in order to explain $g_\mu-2$.
We note that the small ratio $\vev{H_d}/\vev{H}$ suppresses the couplings of $H_d$ to quarks, 
relative to $H$ quark couplings.

The vector-like leptons $X^\ell$, employed in our model to induce the
$\phi\,\bar\ell\ell$ couplings, can be pair-produced in Drell-Yan processes
at the LHC; see Table~\ref{tab:xxsec} for examples of typical cross sections.
However, their discovery signals depend on the dominant branching fractions.  Let us
focus on $X^\mu$ for definiteness, which can decay in a variety of ways.
However, it has only three ``direct" channels  that are not mediated by
mixing: $X^\mu \to \phi \,\mu$, $X^\mu \to H^0_d \,\mu$, and $X^\mu \to H^\pm
\,\nu_\mu$.  Of these, 
given the assumed relation $\lambda_2^2 \gg \lambda_1^2$,
the latter two channels are expected to be dominant in our underlying model.
Here, $H^0_d$ denotes the neutral scalar from the $H_d$ doublet and $H^\pm$
are the associated charged Higgs states, whose main decay modes are
subject to various assumptions about the parameters of the 2-Higgs doublet
potential (see, for example, Ref.~\cite{Davoudiasl:2014mqa} for a
discussion of $H^\pm$ decays in the context of dark $Z$ models).
The exact exclusion limit on $X$ is model dependent.  However, as a rough
estimate, Ref.~\cite{Aad:2015dha} suggests that $m_X\lsim200$ GeV may already be
excluded by the LHC 8 TeV run.  On the other hand, according to 
Table.~\ref{tab:xxsec}, $m_X\gsim300$ still seems viable, and given that $m_\phi$ 
naturalness prefers a relatively lighter $X$, there may be a chance to observe
$X$ pair production at the LHC Run II.
\begin{table}
	\begin{tabular}{ccccc}
		\hline
		$m_X$ [GeV] &200 &300 &400 &600
		\\\hline\hline
		8 TeV & 33 &5.9 & 1.5 & 0.18
		\\
		13 TeV & 79 &16.7 & 5.1 & 0.82 
		\\\hline
	\end{tabular}
	\caption{\label{tab:xxsec}Cross sections for pair production of $X^\ell$ particle at the LHC (in fb).}
\end{table}

The light scalar $\phi$ may also be an interesting target for low energy
experiments, wherever an intense muon beam is available.  The production of
$\phi$ from a muon beam is set by $\lambda^\mu_S \sim 10^{-3}$, making it a
``$\mu$-philic" scalar analogue of a dark photon coupled to charged particles
via kinetic mixing.  Within our setup, for $m_\phi > 2 m_\mu$ (but below $2
m_\tau$) we can expect a $100\%$ branching fraction for $\phi\to \mu^+\mu^-$.
For $m_\phi < 2 m_\mu$ (but above $\sim 1$ MeV), we may have $\phi\to e^+ e^-$
or $\phi \to \gamma \gamma$.  Without further assumptions it is not clear which
one of these two modes will dominate the low mass $\phi$ decays.  However, if
we assume that the entire mass of the electron is generated by an operator
of the type in \eq{dim5mu}, then one can expect $\lambda^e_S \sim (m_e/m_\mu)
\lambda^\mu_S$.  In that case, $\phi \to e^+e^-$ will be the main decay mode in
this mass range.  

For completeness, we also mention that ``dark"
sector states may typically have $\ord{1}$ couplings to $\phi$.  If such states 
are lighter than $m_\phi/2$, then $\phi
\to \text{``invisible"}$ may be the dominant decay mode of $\phi$.  However, in this case 
we may expect $\vev{\phi} \lsim m_\phi$, so that the dark states do not become heavy and 
can furnish on-shell invisible decay final states.

{\it $H\to \phi\phi$:} 
The SM Higgs could mix with the scalar $\phi$ via the following term:
\begin{equation}
	\kappa \left( \phi^\dagger\phi \right)\left( H^\dagger H \right)
	\label{eq:ffhh}
\end{equation}
($\phi(H)^2$ is not allowed because $\phi$ has dark charge.)
Potentially this would lead to $H\to\phi\phi$ decay.
However, requiring $m_\phi$ to stay in the mass range we consider constrains
$\kappa$ to be $\lsim10^{-5}$.   This value for $\kappa$ is stable under
quantum corrections, because the only way to induce such a coupling would be
through a lepton loop, and for a $\tau$ lepton the Yukawa couplings for $H$
and $\phi$ are of order $\sim10^{-2}$, and so the induced contribution is tiny.
With such a small coupling, $H\to \phi\phi \to 4l$ or $\to$invisible should
be negligible.  (Similar consideration also applies to the $H_d$ doublet,
whose vacuum expectation value could be much smaller than that of the SM Higgs
doublet.)

\section{Summary and Conclusions}

In this work, we considered the possibility that a light ``dark" Higgs $\phi$ from a 
hidden sector can be responsible for the measured 3.5 $\sigma$ deviation of $g_\mu-2$ from 
its SM value.  We explored the mass range $m_\phi \lsim 1$~GeV, which 
provides a counterpart to low-energy ``dark" vector boson models that have been 
similarly invoked to address $g_\mu-2$.  In fact, one can assume that 
our dark Higgs $\phi$ is associated with the mechanism responsible for generating 
dark vector boson masses.    

A direct consequence of our scenario is the possibility of $\phi$ emission in decays that include a 
muon; we briefly discussed $\mu\to e\,\nu\,\bar\nu\, \phi$,
$\mu^-p\to\nu_\mu n\phi$, and $K\to \mu \,\nu \,\phi$ as examples of promising search modes
that may lead to signals in future experiments.  Also, the generic assumption
of CP violating couplings of $\phi$ with muons can lead to interesting values 
of the muon electric dipole moment, perhaps as large as $\sim10^{-23}\ e
\cdot\textrm{cm}$, which could potentially  be measured at a future dedicated
storage ring experiment, though a concrete proposal is not currently at hand.  
Similarly, $d_e$ may be within reach of future experiments.
We also discussed that one may anticipate, within a 
generic parameter space, manifestations of lepton flavor violation in decays 
that include a $\phi$, such as $\tau \to \mu \,\phi$, with
$\phi\to e^+e^-$ or $\mu^+\mu^-$.

If $\vev{\phi}\neq 0$, as generally assumed here, we expect 
a new source of mass for muons (and perhaps other leptons) in our 
low energy model.  While we did not specify the value of $\vev{\phi}$ in our setup, 
we pointed out the interesting possibility that for $\vev{\phi} \sim 100$~GeV, 
all or much of the muon mass may originate from $\phi \mu^+ \mu^-$ couplings that explain 
the muon $g_\mu -2$ anomaly.  Hence, a potential signal of our scenario could be a misalignment of the 125 GeV Higgs 
coupling to muons, which may be observable in $H\to \mu^+ \mu^-$ at the LHC, over the next few years.  

We provided a simple UV completion of our scenario, comprising 
weak scale vector leptons and additional ``dark" singlet and doublet 
Higgs scalars.  The high scale model can lead to interesting additional signals at  
the LHC in its Run II, whose generic features were briefly discussed. 

\section{Acknowledgments}

We would like to thank M.~Pospelov for helpful discussions and for informing us
about his work on light scalar effects.  This work is supported in part by the U.S.
Department of Energy under Grant DE-SC0012704.  The work of C.-Y.C is additionally supported by NSERC, Canada. 
Research at the Perimeter Institute is supported in part by the Government of Canada through NSERC and by the Province of Ontario through MEDT.



\appendix*
\section{Appendix}
\subsection{Analytic Expressions for Eqs.~(\ref{Dela}) and (\ref{d})} 

After integrating over $z$ we find
\begin{flalign}
	\Delta a_\ell=
   &\frac{{\lambda_S^\ell}^2}{8 \pi ^2}
   \Bigg[ \frac{3}{2}-r^2-r^2\left(3-r^2\right)\log r
	   \nonumber\\&\qquad\qquad
   -\left(1-r^2\right) \left(4-r^2\right) f(r)\Bigg]
   \nonumber\\
	&+\frac{{\lambda_P^{\ell}}^2}{8 \pi ^2} 
	\Bigg[-\frac{1}{2}-r^2- r^2\left(1-r^2\right) \log r
	   \nonumber\\&\qquad\qquad
		+r^2\left(3-r^2\right) f(r) \Bigg]
\end{flalign}
and
\begin{flalign}
	d_\ell=\frac{\lambda_S^\ell\lambda_P^\ell}{4 \pi ^2 }\frac{e}{2m_\ell}
	\left[1-r^2\log r-\left(2-r^2\right) f(r)\right]
		\label{eq:edm}
\end{flalign}
where
\begin{flalign}
f(r)=\left\{
  \begin{array}{lr}
	  \cos^{-1}\left(\frac{r}{2}\right)\left(4r^{-2}-1\right)^{-\frac{1}{2}} & : r < 2\\[3pt]
	  1 & : r = 2\\
	  \cosh^{-1}\left(\frac{r}{2}\right)
\left(1-4r^{-2}\right)^{-\frac{1}{2}} & : r  > 2.
  \end{array}
\right.
\end{flalign}
Note that for small $r$, $f(r)\approx \pi r/4$.  

\subsection{Flavor Symmetry}

Here, we will present a simple realization of the flavor symmetry that
leads to \eq{L1}, largely as an illustrative example.  Let us consider three separate parities $Z_2^{\ell}$, $\ell=e,\mu,\tau$, broken by
$\vev{S^\ell}$, where $S^\ell$ is a scalar that is $Z_2^{\ell}$ odd.  We do not specify the underlying dynamics for 
$S^\ell$ condensation, since we are only interested in depicting the general symmetry structure.  If desired, that 
physics  can be straightforwardly added to the high energy theory.   
Here, $X^\ell$, as well as SM leptons $L^\ell$ and $\ell_R$ are
all assumed odd under their respective parity.  The usual Yukawa coupling for charged leptons
$y_\ell H \bar L^\ell \ell_R$ can be written down under our assumptions and it will be diagonal in
flavor.  The first term in \eq{L1} can be written as a result of spontaneous symmetry breaking with
\beq
m_X^{\ell \ell'} = \frac{\vev{S^\ell}\vev{S^{\ell'}}}{M}\,,
\label{mll'}
\eeq
where $M$ is a high mass scale.  This is the only source of flavor violation in our setup.  In order to write down a
generic neutrino mass matrix, let us introduce three right-handed neutrinos
$\nu^a_R$, with $a=1,2,3$, that are neutral under $Z_2^\ell$ (assuming three massive neutrinos).
Then, we can have the neutrino mass matrix
\beq
\frac{S^\ell}{M} H \bar L^\ell \nu^a_R + \mbox{\small \text{H.C.}}\,.
\label{numass}
\eeq
For $m_\nu \sim 0.1$~eV and $M$ at the Planck scale $M_P \sim 10^{19}$~GeV, we then find
$\vev{S^\ell}\sim 10^7$~GeV.  This implies that $m_X^{\ell \ell'} \sim 10$~keV, which
can have the right order of magnitude, given the constraints from flavor violating decay bounds
on $\ell \to \ell' \phi$ (see the text for further details).



\end{document}